\documentclass[longauth]{aa}  

\usepackage{graphicx}
\usepackage{txfonts}
\usepackage{soul}
\newcommand{\ktwo}{{\it K2}}
\newcommand{\kepler}{{\it Kepler}}
\newcommand{\cheops}{{CHEOPS}}
\newcommand{\tess}{{\it TESS}}

\newcommand{\target}{{\object{HD~139139}}}

\newcommand{\orcidicon}[1]{
  \href{https://orcid.org/#1}{\includegraphics[scale=0.16]{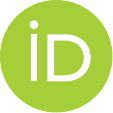}}
}

\begin{document}

   \title{No random transits in CHEOPS observations of \target\ \thanks{This study uses \cheops\ data obtained as part of the Guaranteed Time Observation (GTO) programme CH\_PR110046. Raw and de-trended light curves used in this work are available at the CDS via anonymous ftp to cdsarc.u-strasbg.fr (130.79.128.5) or via \url{http://cdsarc.u-strasbg.fr/viz-bin/qcat?/A+A/}.}}

\author{R. Alonso\inst{1,2}\orcidicon{0000-0001-8462-8126} \and 
S. Hoyer\inst{3}\orcidicon{0000-0003-3477-2466} \and 
M. Deleuil\inst{3}\orcidicon{0000-0001-6036-0225} \and 
A. E. Simon\inst{4}\orcidicon{0000-0001-9773-2600} \and 
M. Beck\inst{5}\orcidicon{0000-0003-3926-0275} \and 
W. Benz\inst{4,6}\orcidicon{0000-0001-7896-6479} \and 
H.-G. Florén\inst{7} \and 
P. Guterman\inst{3,8} \and 
L. Borsato\inst{9}\orcidicon{0000-0003-0066-9268} \and 
A. Brandeker\inst{7}\orcidicon{0000-0002-7201-7536} \and 
D. Gandolfi\inst{10}\orcidicon{0000-0001-8627-9628} \and 
T. G. Wilson\inst{11}\orcidicon{0000-0001-8749-1962} \and 
T. Zingales\inst{12,9}\orcidicon{0000-0001-6880-5356} \and 
Y. Alibert\inst{6,4}\orcidicon{0000-0002-4644-8818} \and 
G. Anglada\inst{13,14}\orcidicon{0000-0002-3645-5977} \and 
T. Bárczy\inst{15}\orcidicon{0000-0002-7822-4413} \and 
D. Barrado Navascues\inst{16}\orcidicon{0000-0002-5971-9242} \and 
S. C. C. Barros\inst{17,18}\orcidicon{0000-0003-2434-3625} \and 
W. Baumjohann\inst{19}\orcidicon{0000-0001-6271-0110} \and 
T. Beck\inst{4} \and 
N. Billot\inst{20}\orcidicon{0000-0003-3429-3836} \and 
X. Bonfils\inst{21}\orcidicon{0000-0001-9003-8894} \and 
Ch. Broeg\inst{4,6}\orcidicon{0000-0001-5132-2614} \and 
S. Charnoz\inst{22}\orcidicon{0000-0002-7442-491X} \and 
A. Collier Cameron\inst{11}\orcidicon{0000-0002-8863-7828} \and 
C. Corral van Damme\inst{23} \and 
Sz. Csizmadia\inst{24}\orcidicon{0000-0001-6803-9698} \and 
P. E. Cubillos\inst{25,19} \and 
M. B. Davies\inst{26}\orcidicon{0000-0001-6080-1190} \and 
A. Deline\inst{27} \and 
L. Delrez\inst{28,29}\orcidicon{0000-0001-6108-4808} \and 
O. D. S. Demangeon\inst{17,18}\orcidicon{0000-0001-7918-0355} \and 
B.-O. Demory\inst{6,4}\orcidicon{0000-0002-9355-5165} \and 
D. Ehrenreich\inst{27,30}\orcidicon{0000-0001-9704-5405} \and 
A. Erikson\inst{24} \and 
A. Fortier\inst{4,6}\orcidicon{0000-0001-8450-3374} \and 
L. Fossati\inst{19}\orcidicon{0000-0003-4426-9530} \and 
M. Fridlund\inst{31,32}\orcidicon{0000-0002-0855-8426} \and 
M. Gillon\inst{28}\orcidicon{0000-0003-1462-7739} \and 
M. Güdel\inst{33} \and 
M. N. Günther\inst{23}\orcidicon{0000-0002-3164-9086} \and 
A. Heitzmann\inst{27} \and 
Ch. Helling\inst{19,34} \and 
K. G. Isaak\inst{35}\orcidicon{0000-0001-8585-1717} \and 
L. L. Kiss\inst{36,37} \and 
E. Kopp\inst{38} \and 
K. W. F. Lam\inst{24}\orcidicon{0000-0002-9910-6088} \and 
J. Laskar\inst{39}\orcidicon{0000-0003-2634-789X} \and 
A. Lecavelier des Etangs\inst{40}\orcidicon{0000-0002-5637-5253} \and 
M. Lendl\inst{20}\orcidicon{0000-0001-9699-1459} \and 
D. Magrin\inst{9}\orcidicon{0000-0003-0312-313X} \and 
P. F. L. Maxted\inst{41}\orcidicon{0000-0003-3794-1317} \and 
Ch. Mordasini\inst{4,6} \and 
V. Nascimbeni\inst{9}\orcidicon{0000-0001-9770-1214} \and 
G. Olofsson\inst{7}\orcidicon{0000-0003-3747-7120} \and 
R. Ottensamer\inst{33} \and 
I. Pagano\inst{42}\orcidicon{0000-0001-9573-4928} \and 
E. Pallé\inst{1,2}\orcidicon{0000-0003-0987-1593} \and 
G. Peter\inst{43}\orcidicon{0000-0001-6101-2513} \and 
G. Piotto\inst{9,12}\orcidicon{0000-0002-9937-6387} \and 
D. Pollacco\inst{44} \and 
D. Queloz\inst{45,46}\orcidicon{0000-0002-3012-0316} \and 
R. Ragazzoni\inst{9,12}\orcidicon{0000-0002-7697-5555} \and 
N. Rando\inst{23} \and 
H. Rauer\inst{24,47,48}\orcidicon{0000-0002-6510-1828} \and 
I. Ribas\inst{13,14}\orcidicon{0000-0002-6689-0312} \and 
N. C. Santos\inst{17,18}\orcidicon{0000-0003-4422-2919} \and 
G. Scandariato\inst{42}\orcidicon{0000-0003-2029-0626} \and 
D. Ségransan\inst{27}\orcidicon{0000-0003-2355-8034} \and 
A. M. S. Smith\inst{24}\orcidicon{0000-0002-2386-4341} \and 
S. G. Sousa\inst{17}\orcidicon{0000-0001-9047-2965} \and 
M. Stalport\inst{49} \and 
Gy. M. Szabó\inst{50,51}\orcidicon{0000-0002-0606-7930} \and 
N. Thomas\inst{4} \and 
S. Udry\inst{20}\orcidicon{0000-0001-7576-6236} \and 
B. Ulmer\inst{43} \and 
V. Van Grootel\inst{29}\orcidicon{0000-0003-2144-4316} \and 
J. Venturini\inst{20}\orcidicon{0000-0001-9527-2903} \and 
F. Verrecchia\inst{52,53} \and 
N. A. Walton\inst{54}\orcidicon{0000-0003-3983-8778}
}

\institute{\label{inst:1} Instituto de Astrof\'\i sica de Canarias, V\'\i a L\'actea s/n, 38200 La Laguna, Tenerife, Spain \\
e-mail: \texttt{ras@iac.es} \and
\label{inst:2} Departamento de Astrof\'\i sica, Universidad de La Laguna, Astrofísico Francisco Sanchez s/n, 38206 La Laguna, Tenerife, Spain \and
\label{inst:3} Aix Marseille Univ, CNRS, CNES, LAM, 38 rue Frédéric Joliot-Curie, 13388 Marseille, France \and
\label{inst:4} Physikalisches Institut, University of Bern, Gesellschaftsstrasse 6, 3012 Bern, Switzerland \and
\label{inst:5} Observatoire Astronomique de l'Université de Genève, Chemin Pegasi 51, CH-1290 Versoix, Switzerland \and
\label{inst:6} Center for Space and Habitability, University of Bern, Gesellschaftsstrasse 6, 3012 Bern, Switzerland \and
\label{inst:7} Department of Astronomy, Stockholm University, AlbaNova University Center, 10691 Stockholm, Sweden \and
\label{inst:8} Division Technique INSU, CS20330, 83507 La Seyne sur Mer cedex, France \and
\label{inst:9} INAF, Osservatorio Astronomico di Padova, Vicolo dell'Osservatorio 5, 35122 Padova, Italy \and
\label{inst:10} Dipartimento di Fisica, Universita degli Studi di Torino, via Pietro Giuria 1, I-10125, Torino, Italy \and
\label{inst:11} Centre for Exoplanet Science, SUPA School of Physics and Astronomy, University of St Andrews, North Haugh, St Andrews KY16 9SS, UK \and
\label{inst:12} Dipartimento di Fisica e Astronomia "Galileo Galilei", Universita degli Studi di Padova, Vicolo dell'Osservatorio 3, 35122 Padova, Italy \and
\label{inst:13} Institut de Ciencies de l'Espai (ICE, CSIC), Campus UAB, Can Magrans s/n, 08193 Bellaterra, Spain \and
\label{inst:14} Institut d’Estudis Espacials de Catalunya (IEEC), Gran Capità 2-4, 08034 Barcelona, Spain \and
\label{inst:15} Admatis, 5. Kandó Kálmán Street, 3534 Miskolc, Hungary \and
\label{inst:16} Depto. de Astrofisica, Centro de Astrobiologia (CSIC-INTA), ESAC campus, 28692 Villanueva de la Cañada (Madrid), Spain \and
\label{inst:17} Instituto de Astrofisica e Ciencias do Espaco, Universidade do Porto, CAUP, Rua das Estrelas, 4150-762 Porto, Portugal \and
\label{inst:18} Departamento de Fisica e Astronomia, Faculdade de Ciencias, Universidade do Porto, Rua do Campo Alegre, 4169-007 Porto, Portugal \and
\label{inst:19} Space Research Institute, Austrian Academy of Sciences, Schmiedlstrasse 6, A-8042 Graz, Austria \and
\label{inst:20} Observatoire astronomique de l'Université de Genève, Chemin Pegasi 51, 1290 Versoix, Switzerland \and
\label{inst:21} Université Grenoble Alpes, CNRS, IPAG, 38000 Grenoble, France \and
\label{inst:22} Université de Paris Cité, Institut de physique du globe de Paris, CNRS, 1 Rue Jussieu, F-75005 Paris, France \and
\label{inst:23} European Space Agency (ESA), European Space Research and Technology Centre (ESTEC), Keplerlaan 1, 2201 AZ Noordwijk, The Netherlands \and
\label{inst:24} Institute of Planetary Research, German Aerospace Center (DLR), Rutherfordstrasse 2, 12489 Berlin, Germany \and
\label{inst:25} INAF, Osservatorio Astrofisico di Torino, Via Osservatorio, 20, I-10025 Pino Torinese To, Italy \and
\label{inst:26} Centre for Mathematical Sciences, Lund University, Box 118, 221 00 Lund, Sweden \and
\label{inst:27} Observatoire Astronomique de l'Université de Genève, Chemin Pegasi 51, 1290 Versoix, Switzerland \and
\label{inst:28} Astrobiology Research Unit, Université de Liège, Allée du 6 Août 19C, B-4000 Liège, Belgium \and
\label{inst:29} Space sciences, Technologies and Astrophysics Research (STAR) Institute, Université de Liège, Allée du 6 Août 19C, 4000 Liège, Belgium \and
\label{inst:30} Centre Vie dans l’Univers, Faculté des sciences, Université de Genève, Quai Ernest-Ansermet 30, 1211 Genève 4, Switzerland \and
\label{inst:31} Leiden Observatory, University of Leiden, PO Box 9513, 2300 RA Leiden, The Netherlands \and
\label{inst:32} Department of Space, Earth and Environment, Chalmers University of Technology, Onsala Space Observatory, 439 92 Onsala, Sweden \and
\label{inst:33} Department of Astrophysics, University of Vienna, Türkenschanzstrasse 17, 1180 Vienna, Austria \and
\label{inst:34} Institute for Theoretical Physics and Computational Physics, Graz University of Technology, Petersgasse 16, 8010 Graz, Austria \and
\label{inst:35} Science and Operations Department - Science Division (SCI-SC), Directorate of Science, European Space Agency (ESA), European Space Research and Technology Centre (ESTEC), Keplerlaan 1, 2201-AZ Noordwijk, The Netherlands \and
\label{inst:36} Konkoly Observatory, Research Centre for Astronomy and Earth Sciences, 1121 Budapest, Konkoly Thege Miklós út 15-17, Hungary \and
\label{inst:37} ELTE E\"otv\"os Lor\'and University, Institute of Physics, P\'azm\'any P\'eter s\'et\'any 1/A, 1117 Budapest, Hungary \and
\label{inst:38} German Aerospace Center (DLR), Institute of Optical Sensor Systems, Rutherfordstraße 2, 12489 Berlin \and
\label{inst:39} IMCCE, UMR8028 CNRS, Observatoire de Paris, PSL Univ., Sorbonne Univ., 77 av. Denfert-Rochereau, 75014 Paris, France \and
\label{inst:40} Institut d'astrophysique de Paris, UMR7095 CNRS, Université Pierre \& Marie Curie, 98bis blvd. Arago, 75014 Paris, France \and
\label{inst:41} Astrophysics Group, Lennard Jones Building, Keele University, Staffordshire, ST5 5BG, United Kingdom \and
\label{inst:42} INAF, Osservatorio Astrofisico di Catania, Via S. Sofia 78, 95123 Catania, Italy \and
\label{inst:43} Institute of Optical Sensor Systems, German Aerospace Center (DLR), Rutherfordstrasse 2, 12489 Berlin, Germany \and
\label{inst:44} Department of Physics, University of Warwick, Gibbet Hill Road, Coventry CV4 7AL, United Kingdom \and
\label{inst:45} ETH Zurich, Department of Physics, Wolfgang-Pauli-Strasse 2, CH-8093 Zurich, Switzerland \and
\label{inst:46} Cavendish Laboratory, JJ Thomson Avenue, Cambridge CB3 0HE, UK \and
\label{inst:47} Zentrum für Astronomie und Astrophysik, Technische Universität Berlin, Hardenbergstr. 36, D-10623 Berlin, Germany \and
\label{inst:48} Institut fuer Geologische Wissenschaften, Freie Universitaet Berlin, Maltheserstrasse 74-100,12249 Berlin, Germany \and
\label{inst:49} Université de Liège, Allée du 6 Août 19C, 4000 Liège, Belgium \and
\label{inst:50} ELTE E\"otv\"os Lor\'and University, Gothard Astrophysical Observatory, 9700 Szombathely, Szent Imre h. u. 112, Hungary \and
\label{inst:51} HUN-REN--ELTE Exoplanet Research Group, Szent Imre h. u. 112., Szombathely, H-9700, Hungary \and
\label{inst:52} Space Science Data Center, ASI, via del Politecnico snc, 00133 Roma, Italy \and
\label{inst:53} INAF, Osservatorio Astronomico di Roma, via Frascati 33, 00078 Monte Porzio Catone (RM), Italy \and
\label{inst:54} Institute of Astronomy, University of Cambridge, Madingley Road, Cambridge, CB3 0HA, United Kingdom
}

\authorrunning{R. Alonso et al.}
   \date{Received July 28, 2023; accepted October 13, 2023}

  \abstract
   {The star \target\ (a.k.a. `the Random Transiter') is a star that exhibited enigmatic transit-like features with no apparent periodicity in \ktwo\ data. The shallow depth of the events ($\sim$200~ppm  -- equivalent to transiting objects with radii of $\sim$1.5~R$_\oplus$ in front of a Sun-like star) and their non-periodicity constitute a challenge for the photometric follow-up of this star.}
   {The goal of this study is to confirm with independent measurements the presence of shallow, non-periodic transit-like features on this object.}
   {We performed observations with \cheops \ for a total accumulated time of 12.75~d, distributed in visits of roughly 20~h in two observing campaigns in years 2021 and 2022. The precision of the data is sufficient to detect 150~ppm features with durations longer than 1.5~h. We used the duration and times of the events seen in the \ktwo\ curve to estimate how many events should have been detected in our campaigns, under the assumption that their behaviour during the \cheops\ observations would be the same as in the \ktwo\ data of 2017. }
   {We do not detect events with depths larger than 150~ppm in our data set. If the frequency, depth, and duration of the events were the same as in the \ktwo\ campaign, we estimate the probability of having missed all events due to our limited observing window would be 4.8\%.}
   {We suggest three different scenarios to explain our results: 1) Our observing window was not long enough, and the events were missed with the estimated 4.8\% probability. 2) The events recorded in the \ktwo\ observations were time critical, and the mechanism producing them was either not active in the 2021 and 2022 campaigns or created shallower events under our detectability level. 3) The enigmatic events in the \ktwo\ data are the result of an unidentified and infrequent instrumental noise in the original data set or its data treatment.}


   \maketitle
%

\section{Introduction}

\begin{table*}
\caption{Observing log of the \cheops\ visits used in this work.}             
\label{table:obs_log}      
\centering          
\begin{tabular}{l l c c c c }     
\hline\hline       
id & \texttt{ObsID}& \texttt{File Key} & Start date & Obs.efficiency& Visit duration\\
\# & & & [UTC]  & [\%]  &  [h] \\
\hline
1 &1461054      & CH\_PR110046\_TG000401\_V0300 & 2021-05-01T11:05:38 & 94.1 & 19.49\\
2 &1490422      & CH\_PR110046\_TG000501\_V0300 & 2021-05-11T07:50:18 & 91.3 & 19.63 \\
3 &1489399      & CH\_PR110046\_TG000601\_V0300 & 2021-05-12T14:21:18 & 91.3 & 19.63 \\
4 &1495676      & CH\_PR110046\_TG000602\_V0300 & 2021-05-17T09:31:17 & 88.6 & 19.43 \\
5 &1506007      & CH\_PR110046\_TG000701\_V0300 & 2021-06-02T19:06:18 & 71.9 & 19.63 \\
6 &1502884      & CH\_PR110046\_TG000603\_V0300 & 2021-06-06T22:24:19 & 72.0 & 19.63 \\
7 &1509091      & CH\_PR110046\_TG000801\_V0300 & 2021-06-08T17:10:18 & 68.5 & 22.91 \\
8 &1512500      & CH\_PR110046\_TG000901\_V0300 & 2021-06-13T18:33:16 & 60.6 & 26.05 \\
9 &1783271      & CH\_PR110046\_TG001001\_V0300 & 2022-04-14T02:10:38 & 71.1 & 19.43\\
10&1792614      & CH\_PR110046\_TG001101\_V0300 & 2022-04-27T12:48:40 & 93.0 & 19.63 \\
11&1791182      & CH\_PR110046\_TG001201\_V0300 & 2022-04-28T22:19:37 & 94.1 & 19.53 \\
12&1797193      & CH\_PR110046\_TG001202\_V0300 & 2022-05-07T12:30:38 & 91.8 & 22.18 \\
13&1795741      & CH\_PR110046\_TG001301\_V0300 & 2022-05-09T05:31:39 & 90.0 & 19.63 \\
14&1804287      & CH\_PR110046\_TG001501\_V0300 & 2022-05-20T16:53:39 & 78.1 & 19.63 \\
15&1808521      & CH\_PR110046\_TG001502\_V0300 & 2022-05-22T10:37:17 & 80.0 & 19.63 \\
 
\hline                    
\hline                  
\end{tabular}
\end{table*}

Thanks to the high precision photometry obtained over hundreds of days with space missions such as \kepler/\ktwo\ (\citealt{Borucki:2010aa,Howell:2014aa}) or \tess\ \citep{Ricker:2015aa}, a few objects have been observed to exhibit behaviours requiring novel explanations (\citealt{Boyajian:2016aa,Rappaport:2019aa}). One of the most intriguing cases, \target\ \citep{Rappaport:2019tb}, was observed during \ktwo\ Campaign 15 (August 23 to November 20, 2017), displaying a series of 28 transit-like events with shallow depths on the order of 200~ppm and no apparent periodicity during the 87 days of continuous observation. The duration and depth of the transits were different at each event, ranging from 0.7 to 8.2 hours and 67 to 408~ppm, respectively. The star shows a 2.7 mag fainter (in G$_{RP}$) companion at an apparent separation of 3.3\arcsec, which is probably physically bound, but it is not clear which star displayed the transit-like events.  Among the different scenarios studied in the discovery paper, none was entirely satisfactory. The possible investigated explanations in \cite{Rappaport:2019tb} included planetary transits in a multiple system, planets orbiting the two stars, a system comprised of only a few planets with huge transit timing variations (TTVs), a disintegrating planet, dust-emitting asteroids, S- and P-type planets around binary systems, dipper-like activity, or multiple short-lived star spots.  \cite{Schneider:2019ac} suggested the possibility of one or a few moons with inclined orbits with respect to the orbital plane of a non-transiting planet. This scenario was excluded with a few radial velocity data points that are stable to the 10~m/s level \citep{Schneider:2019ab}, and in that work, three new scenarios were proposed: a belt of eccentric transiters, transits by interstellar objects, or transits by Solar System objects. Due to its enigmatic nature, \target\ triggered a radio search for technosignatures using the Green Bank Telescope, resulting in a null detection \citep{Brzycki:2019aa}.

Another explored possibility in the original paper was that the \ktwo\ data suffered from some instrumental effect. Thus, a thorough investigation was performed in order to discard potential instrumental effects, including rolling bands, electronic crosstalk, centroid motion analysis during the dips, and background sources. A sophisticated difference image analysis was consistent with a source of the events located close to the target star, but the saturation and bleeding of the columns in the \ktwo\ data complicated the analysis of these images (J. Jenkins, private communication). 

The very shallow depths of the transits and their non-periodicity make any ground-based follow-up currently unfeasible. Due to its location near the ecliptic plane, \target\ has not been observed with \tess\,, and it is not scheduled to be observed until at least Sector 86 (October 2024). With the \ktwo\ mission now over and the high oversubscription factors of larger space telescopes, there has been no independent confirmation of the intriguing events observed with \ktwo\ (`Extraordinary claims require extraordinary evidence', C. Sagan). We describe in this paper our attempt to use the CHaracterising ExOPlanet Satellite \citep[\cheops;][]{Benz:2021aa} to perform such independent confirmation of the transit-like features on \target.

\section{Observations}

CHEOPS is the European Space Agency's first S-class mission, and its payload consisted of an optical telescope with an aperture of 30~cm. This telescope is now orbiting on a Sun-synchronous low-Earth orbit at an altitude of 700~km.  The design of the observing strategy was a challenge due to the duration of the events displayed in the \ktwo\ curve, lasting between 0.7 and 8.2 hours, as well as their intrinsic non-predictability. While a single long visit with a duration of several days is possible using \cheops, it would have a non-negligible impact in the other GTO programmes. Moreover, as our aim is to confirm  the `random transiter' behaviour of \target\ with an independent instrument, detecting a single event with similar characteristics as those observed in the \ktwo\ light curve would suffice for our purpose. If one such event were to be detected at the beginning of a long \cheops\ visit, it would render the observations obtained during the rest of the visit redundant. To avoid this, we adopted the strategy of scheduling one or two weekly visits of 12 continuous orbits each (a duration of about 20~h) at random times when the visibility of the target from \cheops\ was high.  The duration of the individual visits would allow for the detection of individual events, and should one be detected, the programme would be stopped, as its main goal would be reached. In case of non-detection of events, we would schedule additional visits (the scheduling at \cheops\ is planned weekly, in nominal conditions) until a total integrated time that allows for a comparison with the \ktwo\ light curve was reached. This strategy was followed during two campaigns, one in 2021 and another in 2022. The exposure time was 60~s, as recommended for this G=9.56 star. A summary of the observations and their identifiers are presented in Table~\ref{table:obs_log}.  

In total, an accumulated time of 6.93~d was acquired in eight visits of roughly 20~h duration in the 2021 campaign, which spanned 44.4~d. The 2022 campaign observed \target\ in seven additional visits, accumulating a time of 5.82~d distributed over 39.1~d. The time gap between the 2021 and 2022 campaigns, due to the visibility constraints of \cheops, was 303~d.

\section{Analysis}

   \begin{figure*}
   \centering
  \includegraphics[width=\textwidth]{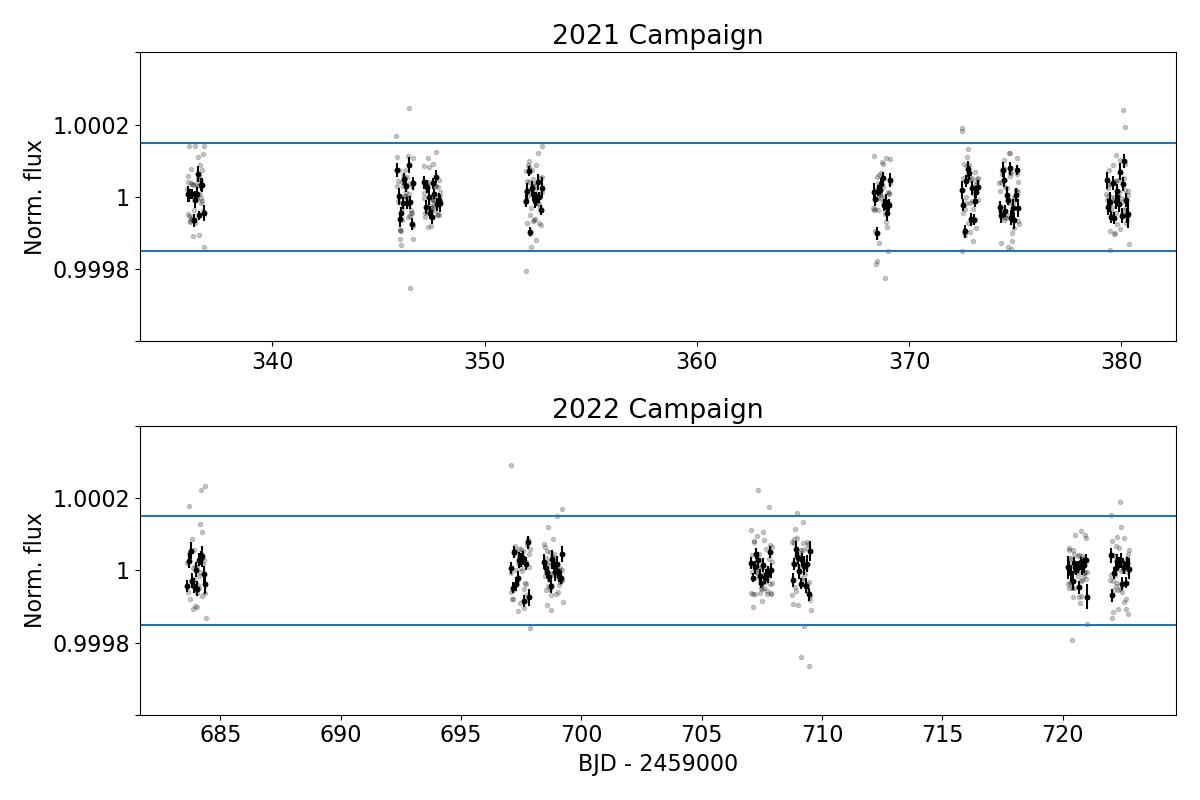}
   \caption{Light curves of \target\ in the two campaigns observed with \cheops. The individual exposures have been de-trended against different vectors using \texttt{pycheops} (see text for details) and binned to 29~mins (grey dots) and one orbit (98.77 min, black dots) for display purposes. The blue horizontal lines represent our detectability limit of 150~ppm, which is smaller than the typical depth of the events detected in \ktwo\ data. No points in the one-orbit binned data lie beyond these lines, which we conservatively set as our detectability limit.}
              \label{fig_two_campaigns}%
    \end{figure*}

The data were processed using the \cheops\ Data Reduction Pipeline (DRP; \citealt{Hoyer:2020aa}), version 14.0. In short, the DRP performs an instrumental calibration (bias, gain, linearisation, and flat-fielding correction) and environmental correction (cosmic ray hits, background, and smearing correction) before extracting the photometric signal of the target in various apertures. After several tests with the different  delivered apertures, we selected the \texttt{R25} (with a radius of 25~px) as the one producing the most precise curves, using the standard deviation of the final curve as a metric. Due to the defocused PSF and the selected radius, the nearby companion of \target\ was also included in the photometric aperture. We performed several attempts of using PSF photometry with the \texttt{PIPE} tool,\footnote{\url{https://github.com/alphapsa/PIPE}} but did not obtain better results than the DRP versions. This could be due to the particular configuration of the system, as the nearby contaminant is partially resolved in the images, and it might be confused with PSF distortions by the pipeline. In the regular \cheops\ operations, the location of the sub-window of the CCD that is downlinked is selected as a result of several monitoring and characterisation (M\&C) visits designed to track the evolution of the instrument. Several factors are taken into account, including PSF shape, dark current, and the location of hot pixels with respect to the PSF. In a few instances, as a result of these observations, the location of the sub-window was changed.\footnote{\url{https://www.cosmos.esa.int/web/cheops-guest-observers-programme/in-orbit-updates}} For our observations, the location of the sub-window in each yearly campaign remained constant, but there was a change between the 2021 and 2022 campaigns of 11~px and 2~px in the X and Y coordinates of the detector, respectively. Despite this change, the noise properties of the data obtained in the two campaigns are comparable.

   \begin{figure*}
   \centering
  \includegraphics[width=\textwidth]{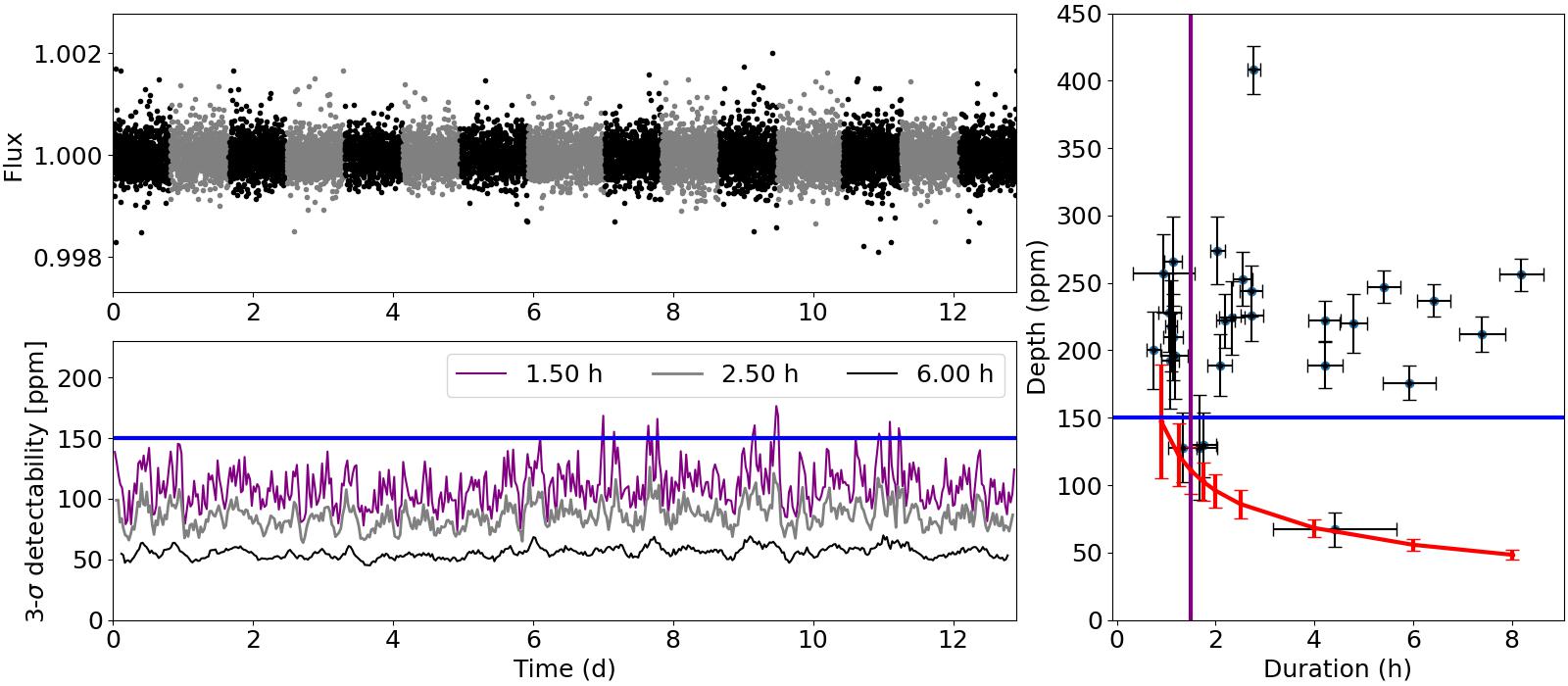}
   \caption{Establishing the transit-like detectability limits of the \cheops\ observations. The upper-left panel shows the 15 \cheops\ visits stitched together with the original 1~min cadence. From this curve, we estimate the 3-$\sigma$ detectability levels for transits of different durations, as shown in the lower-left panel for a sample of three transit durations. The panel on the right shows the duration and depth of the features reported in the original \ktwo\ data (black dots with error bars) and the mean value of the 3-$\sigma$ detectability for different transit durations (red dots and connecting lines). The error bars of the red dots represent the standard deviation of the 3-$\sigma$ detectability curves for each sampled duration. The blue and purple lines represent our selected thresholds for the depth (150~ppm) and duration (1.5~h) of the events that would have been clearly detected in our data. The upper-right quadrant of this panel contains the 16 \ktwo\ events that are used to estimate the observation window of our data and to produce the histogram shown in Figure~\ref{fig:histogram}.  }
              \label{fig_detectability}%
    \end{figure*}

We corrected the DRP data of the remaining systematic effects using the \texttt{pycheops} set of tools \citep{Maxted:2022aa}, version 1.1.0. For the sake of homogeneity, we analysed each visit individually using the same de-correlation vectors for all the visits. The raw curves were $\sigma$ clipped with \texttt{clip\_outliers} by first using a threshold of 4~$\sigma$ and on a second iteration (after removal of the most obvious outliers and re-determination of the mean absolute deviation) by using a more conservative threshold of 6~$\sigma$. We masked the data points with background levels above 0.1~e$^-$, as these were increasing the final dispersion of the curves. Next, the curves were de-correlated against 13 vectors: linear and quadratic terms in time, X and Y positions in the detector, linear terms in background flux, contamination, smear, sin$\phi$,  cos$\phi$, sin$2\phi$, and cos$2\phi$, where $\phi$ is the satellite's orbital phase. As the typical duration of any event in the \ktwo\ light curve was shorter than one \cheops\ visit, which is composed of a minimum of 12 orbits, this correction will have a negligible effect on their detectability. We tested an alternative method to the de-trending described above that models the PSF shape changes using the singular value decomposition of its autocorrelation function, as described in \citet{Wilson:2022aa}, but for this particular case, we did not obtain better results than with the \texttt{pycheops} de-trended curves, an outcome which is probably also related to the effects of the partially resolved nearby contaminant.  

For display purposes, we computed a version in which the data were averaged over one satellite orbit in order to minimise any potential remaining effects at the orbital period. The final light curve is displayed in Figure~\ref{fig_two_campaigns}. The blue horizontal lines in the figure encompass a 150~ppm dimming or brightening. There are no one-orbit binned data points beyond these lines in any of the two campaigns, and we conservatively set this as a first threshold for the detection of transit-like events, with a duration of 1.25 \cheops\ orbits or more ($\sim$2~h), in our data set.

We also constructed versions of the light curves binned to the same cadence as the \ktwo\ data (i.e. $\sim$29~min). As a measure of the obtained precision, the standard deviation of this light curve is 78~ppm, which is higher than the \ktwo\ data by a factor of approximately three. This 29~min cadence light curve is also shown as grey points in Figure~\ref{fig_two_campaigns}. The number of points beyond our threshold of 150~ppm from unity is comparable both below the lower threshold and above the upper threshold, in agreement with the expectations from a constant curve. We also note that there are no consecutive 29~min data points lying beyond our 150~ppm threshold.

Finally, we built a 1~min cadence light curve (i.e. the original cadence), placing the different visits one after the other as if the observations had been performed continuously, with the goal being to better establish the detectability limits. A small gap of 20~mins between the visits was included to mimic the effect of the orbital interruptions. The \texttt{transit\_noise\_plot} tool of \texttt{pycheops} was used to calculate the noise levels producing 1 $\sigma$ detections of transit features for different durations spanning the interval of 0.9 to 8~hours. We used the `scaled' method, in which the transit depth and its standard error are calculated assuming that the true standard errors on the flux measurements are a factor $f$ times the nominal standard error provided by the pipeline. For all the calculated transit durations, this factor was on the order of 1.20 (showing a linear increase from 1.18 for a 0.9~h transit duration to 1.21 for a 8~h duration). The resulting limits at the 3 $\sigma$ level are presented in Figure~\ref{fig_detectability}. Based on these limits, we set two (conservative) constraints to our detectability levels: 150~ppm in transit depth and 1.5~h in total duration.\footnote{There are a few points in the 3-$\sigma$ detectability curve of the 1.5~h transit duration above the 150~ppm threshold, but these take place mostly at the intersections between visits.} With these levels, there are 16 features in the \ktwo\ data set that fall safely above the detectability limits. These features were used to estimate the probability of having missed events due to our observing window.

\section{Comparison to \ktwo\ events}
\label{sec:k2comp}

The light curves of both campaigns of \cheops\ data do not show any clear detection of transit-like features similar to those seen in the \ktwo\ data. In response to this outcome, we sought to establish whether there is a significant probability of this to happen due not observing long enough or because we were unluckily observing at times when no events were happening. Due to the intrinsic non-predictability of the events and their different individual duration and depths, this was a non-trivial task. Our approach was to replicate a simplified version of the \ktwo\ light curve with random starting points for each of the campaigns. The starting points were selected between t$_0$ + $\Delta$t - 87 and t$_0$, where t$_0$ is the time of the first point in one \cheops\ campaign, $\Delta$t is the time span of the observations of one campaign, and 87~d represents the duration of the \ktwo\ observations. With these limits, we were assured that a fraction of the \ktwo\ light curve was sampled entirely in the $\Delta$t of each of the \cheops\ campaigns, which span roughly half the duration of the \ktwo\ data. 

The central times, duration, and depth of the events contain all the needed information from the \ktwo\ light curve for our purpose. For a given starting point, we considered one event as being detected in the \cheops\ data if the number of points (in the 1~min cadence) between the beginning and the end of the event was at least the equivalent to 1.5~h multiplied by the mean observing efficiency of the visits, which is 82.4\%.  We assumed we were only sensitive to the 16 events in \ktwo\ that are both deeper than 150~ppm and longer than 1.5~h, according to the previous results. 

\begin{figure}
   \centering
  \includegraphics[width=\columnwidth]{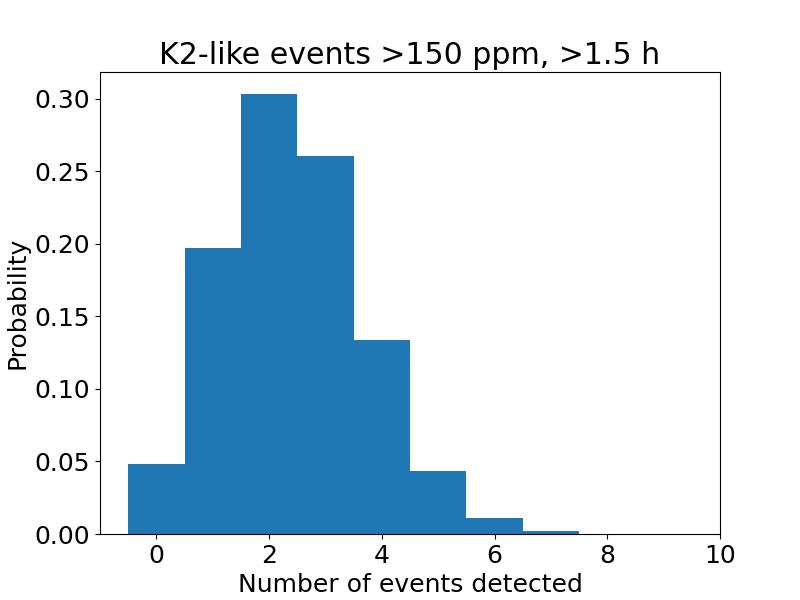}
   \caption{Histogram of the number of events detected in our observation window, under the assumptions that they share the properties of the ones detected in \cite{Rappaport:2019tb}. From the number of times no events were detected, we estimate the probability of having missed all events in our data set due to our limited observing time to be 4.8\%. In this plot, only the \ktwo\ features with depths larger than 150~ppm and durations longer than 1.5~h have been considered.}
              \label{fig:histogram}%
    \end{figure}

We simulated ten thousand curves with random starting points and counted the number of events that were detected with the assumptions above. The result is summarised in the form of a histogram displayed in Figure~\ref{fig:histogram}, and we can estimate the probability to have missed all the events in our observing window as the fraction of times there were no detected events in our simulated curves. Thus, we obtained a probability of 4.8\%.

\section{Summary and conclusions}

We observed \target\ for a total accumulated time that should have allowed for the detection of at least one of the most significant events that was seen in the \ktwo\ data with a reasonable confidence level (95.2~\%). There are a few scenarios that are compatible with these results:

\begin{itemize}

\item Our total observing duration still allows for a small probability (which we estimate to be 4.8\%) of having missed transit-like features with the same characteristics as those in the \ktwo\ data in our window of observations. In the case that these events became less deep or less frequent, as in the epoch of the \ktwo\ observations, the probability of having missed them would obviously increase. This was the main driver for performing the 2022 campaign, that is, to increase the confidence in our non-detectability of \ktwo-like events, which has now reached the subjective level inside the GTO team such that no further observations with \cheops\ will be scheduled. We note that the next possibilities are our preferred explanations.

\item The events seen in the \ktwo\ data were real and `active' at the time of the \ktwo\ observations, and they were `inactive' at the times of \cheops\ observations. There are a few proposed scenarios that could explain this behaviour, such as a cluster of planetesimals or clouds in a shared eccentric orbit that would have transited in 2017 (time of \ktwo\ data). Under this scenario, conducting two \cheops\ campaigns in separate years and not finding any transit-like features in the data could be regarded as the potential physical mechanism having a slight preference for periods of non-activity versus activity.

\item A subset of the previous scenario would be that while \cite{Rappaport:2019tb} made a significant effort to discard possible instrumental effects in the \ktwo\ data, there is still a possibility that the events in the \ktwo\ data set arise from some very infrequent and unidentified noise(s) in that data set. Part of the pixels used in the aperture of the target were exclusively downlinked for the observations of this target (Figure 6 in \citealt{Rappaport:2019tb}), and it was not possible to check for the behaviour of the same pixels in different \kepler/\ktwo\ pointings. This prevents us from checking if one or a few of these pixels could be malfunctioning (e.g. showing a random telegraphic signal, as in \citealt{Hoyer:2020aa}). The fact that \target\ was saturated and bleeding in the \ktwo\ images might have masked such pixels in the core of the stellar image or pixels that were affected by the saturation and bleed.

\end{itemize}

With our observations not providing an independent confirmation of the enigmatic, random transiter configuration of \target, it is improbable that new \cheops\ campaigns would be dedicated to following the behaviour of the system. Nonetheless, should any other independent data provide further support for the anomalous events seen in the \ktwo\ data, we have shown that \cheops\ can reach the precision required for this confirmation.

\begin{acknowledgements}

We thank Saul Rappaport, Andrew Vanderburg and Jon Jenkins for a useful discussion on the instrumental analysis performed on the original paper. 
CHEOPS is an ESA mission in partnership with Switzerland with important contributions to the payload and the ground segment from Austria, Belgium, France, Germany, Hungary, Italy, Portugal, Spain, Sweden, and the United Kingdom. The CHEOPS Consortium would like to gratefully acknowledge the support received by all the agencies, offices, universities, and industries involved. Their flexibility and willingness to explore new approaches were essential to the success of this mission. 
RAl, DBa, EPa, and IRi acknowledge financial support from the Agencia Estatal de Investigación of the Ministerio de Ciencia e Innovación MCIN/AEI/10.13039/501100011033 and the ERDF “A way of making Europe” through projects PID2019-107061GB-C61, PID2019-107061GB-C66, PID2021-125627OB-C31, and PID2021-125627OB-C32, from the Centre of Excellence “Severo Ochoa'' award to the Instituto de Astrofísica de Canarias (CEX2019-000920-S), from the Centre of Excellence “María de Maeztu” award to the Institut de Ciències de l’Espai (CEX2020-001058-M), and from the Generalitat de Catalunya/CERCA programme. 
SH gratefully acknowledges CNES funding through the grant 837319. 
This project was supported by the CNES. 
LBo, TZi, VNa, IPa, GPi, RRa, and GSc acknowledge support from CHEOPS ASI-INAF agreement n. 2019-29-HH.0. 
ABr was supported by the SNSA. 
DG gratefully acknowledges financial support from the CRT foundation under Grant No. 2018.2323 ``Gaseousor rocky? Unveiling the nature of small worlds''. 
TWi and ACCa acknowledge support from STFC consolidated grant numbers ST/R000824/1 and ST/V000861/1, and UKSA grant number ST/R003203/1. 
YAl acknowledges support from the Swiss National Science Foundation (SNSF) under grant 200020\_192038. 
S.C.C.B. acknowledges support from FCT through FCT contracts nr. IF/01312/2014/CP1215/CT0004. 
XB, SC, DG, MF and JL acknowledge their role as ESA-appointed CHEOPS science team members. 
This work has been carried out within the framework of the NCCR PlanetS supported by the Swiss National Science Foundation under grants 51NF40\_182901 and 51NF40\_205606. 
P.E.C. is funded by the Austrian Science Fund (FWF) Erwin Schroedinger Fellowship, program J4595-N. 
The Belgian participation to CHEOPS has been supported by the Belgian Federal Science Policy Office (BELSPO) in the framework of the PRODEX Program, and by the University of Liège through an ARC grant for Concerted Research Actions financed by the Wallonia-Brussels Federation. 
L.D. is an F.R.S.-FNRS Postdoctoral Researcher. 
This work was supported by FCT - Fundação para a Ciência e a Tecnologia through national funds and by FEDER through COMPETE2020 - Programa Operacional Competitividade e Internacionalizacão by these grants: UID/FIS/04434/2019, UIDB/04434/2020, UIDP/04434/2020, PTDC/FIS-AST/32113/2017 \& POCI-01-0145-FEDER- 032113, PTDC/FIS-AST/28953/2017 \& POCI-01-0145-FEDER-028953, PTDC/FIS-AST/28987/2017 \& POCI-01-0145-FEDER-028987, O.D.S.D. is supported in the form of work contract (DL 57/2016/CP1364/CT0004) funded by national funds through FCT. 
B.-O. D. acknowledges support from the Swiss State Secretariat for Education, Research and Innovation (SERI) under contract number MB22.00046. 
This project has received funding from the European Research Council (ERC) under the European Union’s Horizon 2020 research and innovation programme (project {\sc Four Aces}. 
grant agreement No 724427). It has also been carried out in the frame of the National Centre for Competence in Research PlanetS supported by the Swiss National Science Foundation (SNSF). DE acknowledges financial support from the Swiss National Science Foundation for project 200021\_200726. 
MF and CMP gratefully acknowledge the support of the Swedish National Space Agency (DNR 65/19, 174/18). 
M.G. is an F.R.S.-FNRS Senior Research Associate. 
MNG is the ESA CHEOPS Project Scientist and Mission Representative, and as such also responsible for the Guest Observers (GO) Programme. MNG does not relay proprietary information between the GO and Guaranteed Time Observation (GTO) Programmes, and does not decide on the definition and target selection of the GTO Programme. 
CHe acknowledges support from the European Union H2020-MSCA-ITN-2019 under Grant Agreement no. 860470
(CHAMELEON). 
KGI is the ESA CHEOPS Project Scientist and is responsible for the ESA CHEOPS Guest Observers Programme. She does not participate in, or contribute to, the definition of the Guaranteed Time Programme of the CHEOPS mission through which observations described in this paper have been taken, nor to any aspect of target selection for the programme. 
K.W.F.L. was supported by Deutsche Forschungsgemeinschaft grants RA714/14-1 within the DFG Schwerpunkt SPP 1992, Exploring the Diversity of Extrasolar Planets. 
This work was granted access to the HPC resources of MesoPSL financed by the Region Ile de France and the project Equip@Meso (reference ANR-10-EQPX-29-01) of the programme Investissements d'Avenir supervised by the Agence Nationale pour la Recherche. 
ML acknowledges support of the Swiss National Science Foundation under grant number PCEFP2\_194576. 
PM acknowledges support from STFC research grant number ST/M001040/1. 
This work was also partially supported by a grant from the Simons Foundation (PI Queloz, grant number 327127). 
NCSa acknowledges funding by the European Union (ERC, FIERCE, 101052347). Views and opinions expressed are however those of the author(s) only and do not necessarily reflect those of the European Union or the European Research Council. Neither the European Union nor the granting authority can be held responsible for them. 
S.G.S. acknowledge support from FCT through FCT contract nr. CEECIND/00826/2018 and POPH/FSE (EC). 
GyMSz acknowledges the support of the Hungarian National Research, Development and Innovation Office (NKFIH) grant K-125015, a a PRODEX Experiment Agreement No. 4000137122, the Lend\"ulet LP2018-7/2021 grant of the Hungarian Academy of Science and the support of the city of Szombathely. 
V.V.G. is an F.R.S-FNRS Research Associate. 
NAW acknowledges UKSA grant ST/R004838/1.

\end{acknowledgements}

\bibliographystyle{aa} 
   
\bibliography{refs_hd139139.bib} 

\end{document}